# Trion Formation Dynamics in Monolayer Transition Metal Dichalcogenides

Akshay Singh[1], Galan Moody[1], Kha Tran[1], Marie E. Scott[2], Vincent Overbeck[3], Gunnar Berghäuser[3], John Schaibley[2], Edward J. Seifert[1], Dennis Pleskot[4], Nathaniel M. Gabor[4], Jiaqiang Yan[5,6], David G. Mandrus[5,6,7], Marten Richter[3], Ermin Malic[8], Xiaodong Xu[2,9], and Xiaoqin Li[1*]

[1] Department of Physics, University of Texas at Austin, Austin, TX 78712, USA.
[2] Department of Physics, University of Washington, Seattle, Washington 98195, USA.
[3] Institut für Theoretische Physik, Nichtlineare Optik und Quantenelektronik, Technische Universität Berlin, 10623 Berlin, Germany
[4] Department of Physics and Astronomy, University of California, Riverside, California 92521, USA
[5] Materials Science and Technology Division, Oak Ridge National Laboratory, Oak Ridge, Tennessee 37831, USA
[6] Department of Materials Science and Engineering, University of Tennessee, Knoxville, Tennessee 37996, USA
[7] Department of Physics and Astronomy, University of Tennessee, Knoxville, Tennessee 37996, USA
[8] Department of Applied Physics, Chalmers University of Technology, Gothenburg 41258, Sweden
[9] Department of Materials Science and Engineering, University of Washington, Seattle, Washington 98195, USA

*e-mail address: elaineli@physics.utexas.edu

We report charged exciton (trion) formation dynamics in doped monolayer transition metal dichalcogenides, specifically molybdenum diselenide ($MoSe_2$), using resonant two-color pump-probe spectroscopy. When resonantly pumping the exciton transition, trions are generated on a picosecond timescale through exciton-electron interaction. As the pump energy is tuned from the high energy to low energy side of the inhomogeneously broadened exciton resonance, the trion formation time increases by ~ 50%. This feature can be explained by the existence of both localized and delocalized excitons in a disordered potential and suggests the existence of an exciton mobility edge in transition metal dichalcogenides.

**Keywords:** atomically thin semiconductors, excitons, trions, ultrafast dynamics

**PACS:** 78.47.jg, 78.47.D, 78.67.-n, 71.35.Pq

Optical excitation of semiconductors generates electron-hole pairs, called excitons, held together via Coulomb interactions. In the presence of residual free electrons, excitons interact with the surrounding charges, ultimately binding to form charged excitons called trions [1-3]. The ultrafast formation time for these quasiparticles has not been experimentally accessible in atomically thin transition metal dichalcogenides (TMDs). Yet, it is critical for evaluating and improving performance of optoelectronic devices based on this emerging class of materials with many fascinating properties tunable via layer thickness, strain, doping, and stacking [4-12].

Excitons and trions in monolayer TMDs are stable at room temperature due to their remarkably large binding energies in the range of a few hundred meV and tens of meV, respectively [4,5,10,11,13-17]. The exciton to trion formation (ETF) process is energetically favorable, leading to a characteristic trion wavefunction as visualized in Fig. 1(a). In the figure, the positions of a hole and an electron are fixed and chosen to be separated by 1 nm, corresponding approximately to the exciton Bohr radius for this material [18,19]. The probability to find a second electron is calculated to be the highest near the hole due to the attractive Coulomb force [20]. Our calculation takes into account the drastically modified screening in monolayer materials and the substrate effect, leading to a trion binding energy close to that measured experimentally. In the presence of disorder, the momentum of the center of mass motion is no longer a good quantum number as assumed in the calculation. We anticipate the trion formation time to be modified in a disordered potential.

Distinct exciton and trion wavefunctions and properties make ETF dynamics a fundamentally important physical process. First, being charged composite quasiparticles, trions drift in an applied electric field [21]. Thus, ETF modifies photoconductivity and energy transport [12]. Second, ETF is an important exciton population relaxation channel and is therefore critical for interpreting exciton decay dynamics and the relative spectral weight of trions and excitons in photoluminescence [10,22]. Finally, valley dynamics and radiative relaxation are expected to be different for excitons and trions in TMDs [23-25], making the ETF process highly relevant for valleytronics and light emitting devices.

In this work, we investigate ETF dynamics in monolayer MoSe$_2$ using ultrafast, two-color pump-probe spectroscopy with properly chosen spectral and temporal resolutions. When resonantly pumping the exciton and probing the trion transitions, the ETF process is manifested as a finite rise time $\tau_f$ in the differential reflectivity signal, as a function of the delay time between the two pulses, as illustrated in Fig. 1(b) [24,26-28]. The trion formation time $\tau_f$ increases as the pump energy is tuned from the high energy to low energy side of the inhomogeneously broadened exciton resonance. This observation suggests the presence of an effective exciton "mobility edge", i.e. below (above) a certain energy, the center of mass motion of the excitons is localized (delocalized). Our studies articulate the role of disorder, distinguish between coexisting quasiparticles with different characteristics, and provide a more accurate picture of the complex quasiparticle dynamics present in TMDs [23,29,30].

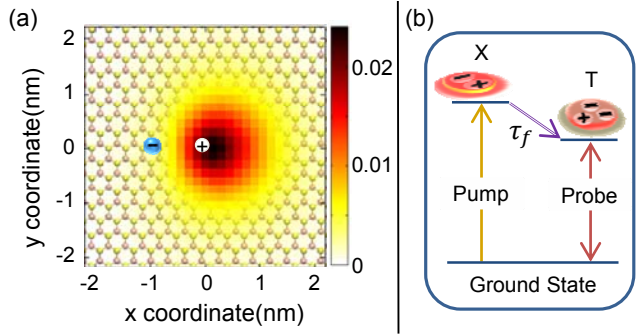

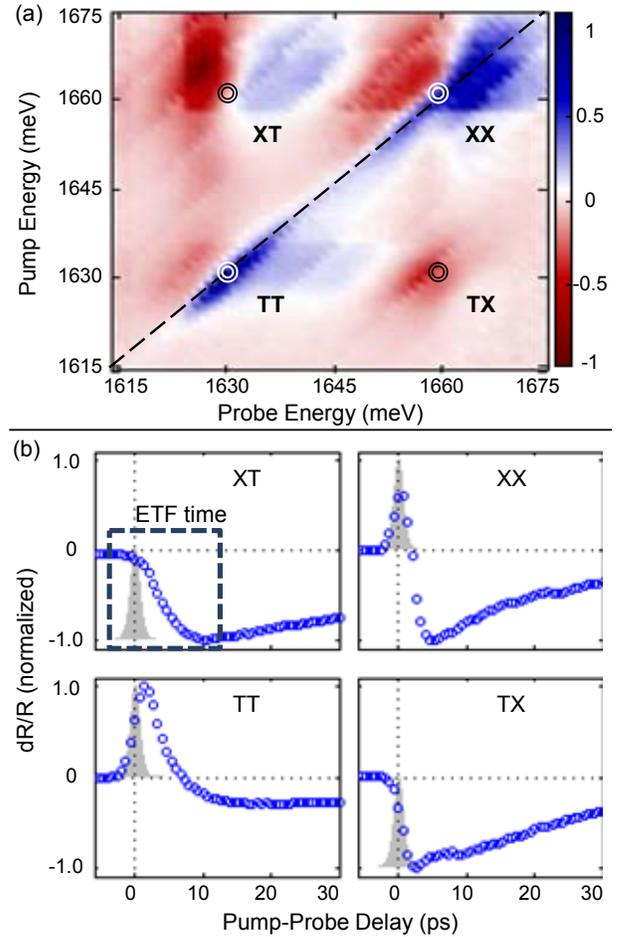

Figure 1: (color online): (a) Calculated relative trion wavefunction. The positions of a hole and an electron (indicated by + and – signs, respectively) are fixed at 1 nm separation corresponding to exciton Bohr radius. The asymmetric distribution of the wavefunction is due to the attractive (repulsive) Coulomb force between the hole (electron) and the second electron. (b) Energy diagram illustrating the two-color pump-probe scheme for measuring the ETF with a finite trion formation time $\tau_f$. Pump and probe energies are tuned to the exciton and trion resonances, respectively.

We study a naturally *n*-doped monolayer MoSe$_2$ mechanically exfoliated on a SiO$_2$/Si substrate and hence we are studying negative trions [10,20]. The sample temperature is held at 13 K for all experiments to reduce phonon interaction induced resonance broadening. The narrow spectral linewidths in combination with large trion binding energy lead to spectrally well-resolved exciton and trion resonances in this high quality sample. The experimental setup for the two-color pump/probe experiment is described in [20]. Briefly, pump and probe beams derived from a Ti:sapphire laser are spectrally filtered independently using grating-based pulse-shapers, generating ~ 0.7 nm (~ 1.5 ps) full-width at half-maximum (FWHM) pulses. The pump and probe beams are recombined and focused collinearly onto the sample with a spot size of ~ 2 $\mu$m. We use cross linearly polarized pump and probe pulses to suppress laser scatter from the pump pulse reaching the detection optics.

Figure 2: (color online): (a) Normalized two-color differential reflectivity (dR/R) measurement at zero pump/probe delay. Exciton (XX) and trion (TT) peaks appear on the diagonal dashed lined. Exciton-trion coupling appears in the spectrum as cross diagonal peaks (XT and TX). The notation XT represents the optical response while resonantly pumping the exciton and probing the trion. Other notations are defined in a similar manner with the first letter referring to the pump energy and the second letter referring to the probe energy. (b) Delay scans for the four peaks indicated by the circles in (a). XT has a finite rise time to the maximal dR/R signal while other peaks have pulse-width limited rise times. The excitation pulse (the gray shaded area) is shown for comparison.

We first utilize spectral scans in the two-color pump-probe experiment to generate a full two-dimensional

(2D) map of the differential reflectivity dR/R = [R–$R_0$]/$R_0$, where R ($R_0$) is the probe reflectivity with (without) the pump present. The diagonal peaks in the 2D map (Fig. 2(a)) are associated with the trion and exciton resonances at 1631 meV and 1662 meV respectively, whereas cross-diagonal peaks (XT, TX) reveal exciton-trion coupling and conversion processes. The delay time (~0.7 ps) is chosen in this manuscript to slightly enhance the visibility of all four peaks in the two-dimensional pump-probe spectrum in Fig. 2a. The energy separation between the trion and exciton (~ 31 meV) agrees well with the trion binding energy from previous studies on monolayer $MoSe_2$ and our calculation [10,20,31]. The line-shapes (absorptive or dispersive) of the different peaks reflect interplay between the relative phase of the reflected probe and nonlinear signal and many-body effects as shown in our previous work [32]. For example, the distinct lineshape of the TX peak is due to coherent coupling between the exciton and trion.

We now examine quasiparticle ultrafast dynamics by taking delay scans (Fig. 2(b)) while pump and probe are tuned to measure each peak in the 2D map. The exact pump/probe energies chosen are indicated by the circles in Fig. 2(a). When the pump and probe energies are resonant with the exciton (XX) and trion (TT) transitions, these quasiparticles form rapidly within our temporal resolution (~ 1 ps) and decay on tens of picoseconds time scales [23,33,34]. The dynamical evolution of dR/R signal is complex and includes a change in sign, which has been attributed to higher order optical processes and/or energy renormalization in previous studies [33,34]. Additionally, the slower decay of the TT signal compared to the XX signal suggests a longer relaxation time for trions consistent with earlier experiments [23]. Furthermore, pumping at the trion resonance and probing at the exciton resonance (TX) also leads to a fast rise in the dR/R signal limited by the temporal resolution in our experiments, which is consistent with instantaneous coherent coupling between the exciton and trion as previously discussed [32,35]. The delay time (~0.7 ps) is chosen to slightly enhance the visibility of all four peaks in the two-dimensional pump-probe spectrum in Fig. 2(a).

In contrast, a finite rise time in dR/R signal beyond the pulse temporal width was observed when pumping at the exciton resonance and probing at the trion resonance (panel XT in Fig. 2(b)). This finite rise time is a signature of the ETF process, which is the focus of this paper [26,36]. We analyze the ETF process in more detail by carefully choosing pump and probe energies. The pump ($\hbar\omega_{pump}$) and probe ($\hbar\omega_{pr}$) energies are shown in Figure 3a overlaid with the degenerate spectral scan.

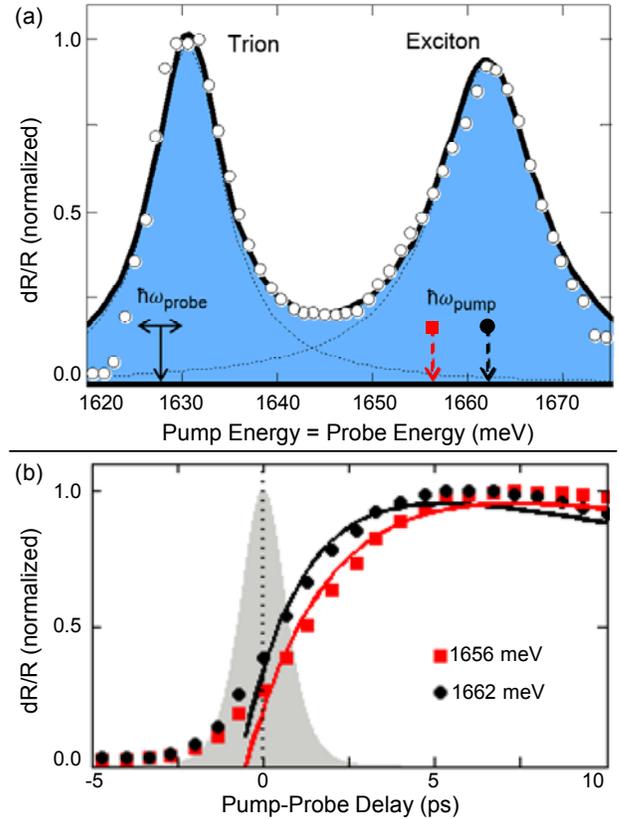

Figure 3: (color online) (a) Normalized degenerate differential reflectivity measurement at zero delay between the pump and probe pulses. Lorentzian fits to spectral peaks are shown with dotted lines (trion: ~1631 meV, FWHM ~ 4.5 meV; exciton: ~1662 meV, FWHM ~ 6.5 meV). The pump ($\hbar\omega_{pump}$) and probe ($\hbar\omega_{pr}$) energies for the non-degenerate ETF experiment are indicated by the arrows. (b) Integrated XT delay scans for the two pump excitation energies (1656 and 1662 meV). When pumping at 1662 meV, the rise to maximum dR/R signal is faster compared to excitation at 1656 meV. The fits to the model (described in text) are shown as lines.

The probe energy is fixed to the lower energy side of the trion (1627 meV) to minimize probing the exciton and trion resonances simultaneously. The differential reflectivity signal is integrated over the probe energy within a ±2 meV window to enhance the signal-to-noise ratio. Pump-probe delay scans focusing on the initial rise dynamics of peak XT for two pump excitation energies are shown in Fig. 3(b).

To quantify the ETF dynamics, we use a simple fitting function that takes in account both the rise and decay components associated with the ETF and trion relaxation processes, respectively,

$$dR/R = A_1[1 - A_2 \exp(-\Delta t / \tau_f)] * \exp(-\Delta t / t_d), \quad (1)$$

where $A_1$ and $A_2$ are fit amplitudes, $\tau_f$ is the trion formation time, and $t_d$ is the trion relaxation time. In using this model, we have assumed instantaneous exciton formation, which is supported by the time evolution of the XX peak shown in Fig. 2(b).

A trion formation time $\tau_f = 1.6\pm0.1$ ps is extracted from the fit to the data taken with the pump tuned to 1662 meV (black dots in Fig. 3(b)). $\tau_f$ for ETF has been previously measured in several classes of materials and varies from tens of picoseconds in GaAs and CdTe quantum-wells [24,36-38] to a few femtoseconds in carbon nanotubes [26,27]. A precise and direct comparison between different material systems is difficult due to different conditions under which experiments are conducted (e.g. doping density and excitation conditions). Nevertheless, an intermediate trion formation time observed here is consistent with the general understanding of the exciton properties in these different classes of materials [39]. In Table I, we compare exciton Bohr radius, trion binding energy and ETF times from three different groups of materials. We observe that the ETF time decreases with increasing trion binding energy, determined by Coulomb interaction which in turn depends on dimensionality as well as screening. Although ETF time depends on several other factors, including excitation power, doping density, temperature, and exciton localization length, the consistent trend in ETF time among these different classes of materials indicates that strength of the screened Coulomb interactions is the key factor that determines the order of magnitude of ETF time. Below, we show that measurements of the ETF time can actually shine light on the nature of exciton states in a disordered potential. It is essential to characterize exciton localization for applications involving quasiparticle transport and photoconductivity.

| Physical System | Exciton Bohr Radius, Dimension | Trion Binding Energy | Trion Formation Time |
|---|---|---|---|
| Carbon Nanotubes | 1 nm, 1D system [40,41] | 60- 130 meV [27,40] | 60- 150 fs [26,27] |
| TMD (MoSe$_2$) | 1 nm, 2D system [18] | 30 meV [10] | 2 ps (this study) |
| Quantum Wells (GaAs, CdTe) | 15 nm, 2D system [42] | 2, 3 meV [24,36,38] | 100 ps, 60 ps) [24,38] |

TABLE I: Comparison of different materials illustrating trion formation time dependence on exciton bohr radius and dimensionality.

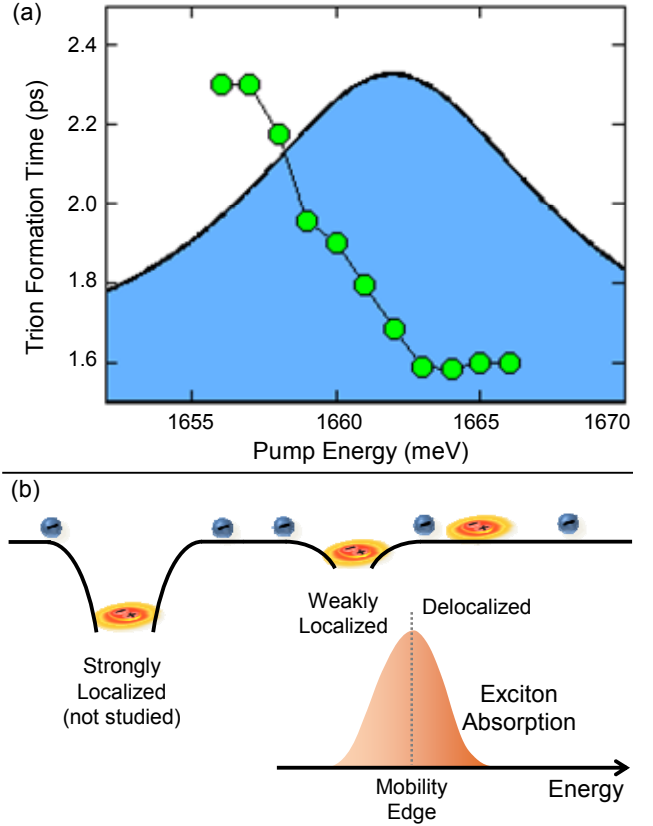

Figure 4 (color online): (a) Dependence of trion formation time on changing pump energy across exciton resonance. The exciton resonance is overlaid for reference. (b) Illustration of different disorder potentials. Excitons below (above) the mobility edge are localized (delocalized) and take longer (shorter) to capture an electron and form a trion.

Interestingly, the trion formation time depends on the exact pump energy under the exciton resonance. This dependence is already observable from the time traces presented in Fig. 3(b) at two different pump energies. By systematically tuning the pump energy from the higher energy to low energy side of the inhomogeneously broadened exciton resonance, we find that the trion formation time increases from 1.6 ps to 2.3 ps, shown in Fig. 4(a)). We attribute the dependence of the trion formation time on excitation energy to localization of the center of mass motion of excitons in the presence of disorder potentials which may be ascribed to impurities, vacancies, or strain from the substrate.

Different types of disorder potentials and their effects on excitons are illustrated in Fig. 4(b). An exciton may be strongly spatially localized via deep potential traps associated with certain type of impurities and sample edges. This type of strongly localized exciton has been investigated in WSe$_2$ recently through single photon emission experiments [43-45]. Typically, these excitons are red-shifted tens of meV below the trion transition. Thus, we do not probe these bound states in our experiments performed under resonant excitation conditions. An exciton may also be weakly localized by shallow potentials. The energy of these weakly localized states is only slightly red-shifted compared to the delocalized states, leading to inhomogeneous broadening of the exciton resonance. A "mobility edge" separates these two types of excitons in energy [37,46-48]. The center-of-mass wavefunction of delocalized excitons with energy above the mobility edge extends across a large spatial region. The large extension increases the probability for this exciton to interact with residual background carriers [37], resulting in a faster trion formation time for high energy excitons, as shown in Fig. 4(a). On the other hand, the wavefunction of a localized exciton with energy below the mobility edge is centered at a particular spatial location and decays away from it. The in-plane localization of the exciton wavefunction reduces the exciton-free carrier interaction and results in a longer formation time. While the concept of a mobility edge is theoretically hypothesized as a sharp energy that separates delocalized and localized states, the transition occurs across a spectral region determined by the sample quality and disorder as observed in our experiments.

The fast ETF times reported here suggest that the ETF process is an efficient exciton relaxation channel that must be considered when interpreting ultrafast dynamics in time-resolved spectroscopy experiments. The ETF time dependence on exciton localization is a particularly interesting result, since it suggests the existence of exciton mobility edge, a concept that has remained unexplored in studies of quasiparticle transport in TMDs so far. While similar concepts have been discussed in conventional quasi-2D quantum wells such as GaAs, the underlying physical mechanisms relevant for these phenomena observed in TMDs are fundamentally different. In high quality GaAs quantum wells, the disorder potential typically arises from well width fluctuations, a mechanism that cannot be invoked in these monolayer semiconductors. Other manifestations of disorder potentials in monolayer TMDs include a direct measurement of inhomogeneous broadening at low temperature [30] and a Stokes shift (a few meV) between exciton resonances measured in PL and absorption (data not included). Future experiments that combine high spectral resolution with atomic or mesoscopic spatial resolution might reveal how different types of impurities and disorder potentials with different characteristic length scales influence optical selection rules and ultrafast quasiparticle dynamics in TMDs [49,50].

**Acknowledgements**: The spectroscopic experiments performed by Singh were supported jointly through ARO W911NF-15-1-0088 and AFOSR FA9550-10-1-0022. The work by Seifert was supported by NSF DMR-1306878. The collaboration on sample preparation between UT-Austin and UC-Riverside was supported as part of the SHINES, an Energy Frontier Research Center funded by the U.S. Department of Energy (DOE), Office of Science, Basic Energy Science (BES) under Award # DE-SC0012670. Tran, Li, Pleskot, and Gabor have received support from the SHINES. The UW team supported by U.S. DoE, BES, Materials Sciences and Engineering Division (DE-SC0008145) prepared samples and contributed to interpretation of spectroscopic data. The samples were provided by Yan and Mandrus at ORNL with support by U.S. DoE, Office of Basic Energy Sciences, Materials Sciences and Engineering Division. Malic acknowledges funding form the EU Graphene Flagship (CNECT-ICT-604391). Li also acknowledges the support from a Humboldt fellowship, which facilitated the collaboration on theoretical studies performed by Overbeck, Berghäuser, Richter, and Malic.